\documentclass{bioinfo}
\pdfoutput=1
\usepackage{booktabs, url}
\usepackage{mathtools}

\begin{document}
\firstpage{1}

\title[Immunophenotypes of AML]{Immunophenotypes of Acute Myeloid Leukemia \\From Flow Cytometry Data Using  Templates}
\author[Azad et al.]{Ariful Azad$^{1}$, Bartek Rajwa\,$^{2}$ and Alex Pothen\,$^3$\footnote{to whom correspondence should be addressed}}
\address{
$^{1}$azad@lbl.gov, Computational Research Division, Lawrence Berkeley National Laboratory, California, USA\\
$^{2}$brajwa@purdue.edu, Bindley Bioscience Center, Purdue University, West Lafayette, Indiana, USA \\
$^{3}$apothen@purdue.edu, Department of Computer Science, Purdue University, West Lafayette, Indiana, USA
}



\maketitle

\begin{abstract}
\section{Motivation:}
We investigate whether a template-based classification pipeline could be used to 
identify immunophenotypes in (and thereby classify) a  heterogeneous disease with many
subtypes. The disease we consider here is Acute Myeloid Leukemia, 
which is heterogeneous at the morphologic, cytogenetic and molecular levels, with 
several known subtypes. The prognosis and treatment for AML depends on the subtype. 
\section{Results:}
We apply flowMatch, an algorithmic pipeline for flow cytometry data 
created in earlier work, 
to compute templates succinctly summarizing classes of AML and healthy samples. 
We develop a scoring  function that accounts for features of the AML data such as 
heterogeneity to identify immunophenotypes  corresponding to various AML subtypes, 
including APL. 
All of the AML samples in the test set are classified correctly with high confidence. 

\section{Availability:}
flowMatch is available  at
{\tt www.bioconductor.org/}
\newline {\tt packages/devel/bioc/html/flowMatch.html}; programs
\newline  specific to immunophenotyping  AML are at
{\tt www.cs.purdue.edu/}
\newline {\tt homes/aazad/software.html}.

\section{Contact:} \href{apothen@purdue.edu}{apothen@purdue.edu}
\end{abstract}

\section{Introduction}

Can Acute Myeloid Leukemia (AML) samples be distinguished  from healthy ones  using 
flow cytometry data  from blood or bone marrow with a template-based classification method? 
This method builds a template for each class  to summarize the  samples belonging to the class,
and uses them to classify new samples. 
This question is interesting because AML is a heterogeneous disease with several subtypes
and hence it is not clear that a template can succinctly  describe all types  of AML. 
Furthermore, we wish to identify
immunophenotypes (cell types in the bone marrow and blood) 
that are known to be characteristic of  subtypes of AML. 
Pathologists use these immunophenotypes to visualize AML and its subtypes,
and a computational procedure that can provide this information would be 
more helpful in clinical practice than a classification score that 
indicates if an individual is healthy or has AML. 

In earlier work, we have developed a template-based classification method for 
analyzing flow cytometry (FC) data,
which consists of measurements of morphology (from scattering) and the expression of 
multiple biomarkers (from fluorescence) at the single-cell level. 
Each FC sample consists of  hundreds of thousands or more of such single-cell measurements,
and a study could consist of thousands of samples from different individuals
at different time points under different experimental conditions~\citep{aghaeepour2013critical,shapiro2005practical}. 
We have developed an algorithmic pipeline for various steps in   processing this 
data~\citep{azad2013classifying,azad2010identifying,azad2012matching}.  
We  summarize each sample by means of the cell populations that it contains.
(These terms are defined in Table~\ref{tab:terminology} 
and illustrated in Fig.~\ref{fig:template_example}.)
Similar samples belonging to the same class are  described 
by a template for the class. A  template consists of meta-clusters that 
characterize the cell populations present in  the samples that constitute the class. 
We compute templates from the samples, and organize the templates into a template tree. 
Given a sample to classify, we compare it with the nodes in the template tree,
and classify it to the template that it is closest to. 
A combinatorial measure for the dissimilarity of two samples or two templates,  computed 
by means of a mixed edge cover in a graph model (described in the next section),
is at the heart of  this approach. 

We have applied our algorithmic pipeline for template-based classification to  various problems:
to distinguish the phosphorylation state of T cells; 
to study the biological, temporal, and technical variability of cell types in the 
blood of healthy individuals; to characterize changes in the immune cells of 
Multiple Sclerosis patients undergoing drug treatments; 
and to predict the vaccination status of HIV patients. 
However, it is not clear if the AML data set can be successfully analyzed with 
this scheme, 
since AML is a hetereogeneous disease at the morphologic, cytogenetic and molecular levels,
and a few  templates may not describe all of its  subtypes. 

AML is a disease of myeloid stem cells that differentiate to form several types of cells 
in the blood and marrow. It is characterized by the profusion of immature myeloid cells,
which are usually prevented from maturing due to the disease. 
The myeloid 
stem cell differentiates in several steps to form myeloblasts
and other cell types in a hierarchical process. 
This hierarchical differentiation process could be blocked at different cell types, 
leading to the multiple  subtypes of AML. 
Eight different subtypes of AML based on cell lineage are included in the French-American-British 
Cooperative Group (FAB)  classification scheme~\citep{bennett1985proposed}. 
(A different World Health Organization (WHO)  classification scheme has also 
been published.)
Since the prognosis  and treatment varies greatly among the subtypes of AML, 
accurate diagnosis is critical. 

We extend our earlier  work on template-based classification here by developing 
a scoring function that accounts for the subtleties of 
FC data of AML samples. Only a small number of the myeloid cell populations in AML samples
are specific to AML, and there are a larger number of cell populations that these samples
share with healthy samples. Furthermore, the scoring function needs to account for the diversity
of the myeloid cell populations in the various subtypes of AML. 

Our work has the advantage of identifying immunophenotypes of clinical interest in AML 
from the templates. 
Earlier work on the AML dataset we work with has classified AML samples using 
methods such as nearest neighbor classification, logistic regression, 
matrix relevance learning vector quantization, etc.,
but they have not identified these 
immunophenotypes; e.g.,~\citep{biehl2013analysis,manninen2013leukemia,qiu2012inferring}. 

Template-based classification has the advantage of being more robust than nearest neighbor
classification since a template summarizes the characteristic properties of a class
while ignoring small sample-to-sample variations. 
It is also  scalable to large numbers of samples, since we compare a sample to be classified
only against a small number of templates rather than the much larger number of samples. 
The comparisons with the templates can  be performed efficiently using 
the structure of the template tree. 
It also reduces the data size by clustering the data to identify cell populations
and then working with the statistical distributions characterizing the cell populations,
in contrast to some of the earlier approaches that work with data sets 
even larger than the FC data 
by creating  multiple variables from a marker 
(reciprocal, powers, products and quotients of subsets of the markers, etc.). 

\begin{table}[!tbp]
  \centering
\processtable{Summary of terminology 
 \label{tab:terminology}}
{\begin{tabular}{p{0.22\linewidth} p{0.70\linewidth}}
      \toprule
      Terms    & Meaning \\
     \toprule
      Sample       & From FC data.  Characterized by the  collection of cell populations included 
                    within it.   \\
            \midrule
       Cell population (cluster) & A group of cells with identical morphology and expressing similar biomarkers, e.g., helper T cells, B cells. Computed from a sample. \\
            \midrule

      Meta-cluster       & A set of similar cell clusters from different samples. Computed from similar clusters in  samples.   \\
            \midrule
      Template & A  collection of meta-clusters from samples of the same class.   \\
      \bottomrule
\end{tabular}}{}
\vspace{-.2in}
\end{table}

Template-based classification has been employed in other areas such as character, face, and image recognition, 
but its application to FC is relatively recent. 
In addition to our work, 
templates have been used for 
detecting the effects of phosphorylation~\citep{pyne2009automated},  
evaluating the efficiency of data transformations~\citep{finak2010optimizing},  
and labeling clusters across samples~\citep{spidlen2013genepattern}.   

\vspace{-.2in}

\section{Methods}


\subsection{The AML Dataset}
We have used an FC dataset on AML that  was included in the 
DREAM6/FlowCAP2 challenge of  2011. 
The dataset consists of FC measurements of peripheral blood or bone marrow aspirate collected from 
$43$ AML positive patients and $316$ healthy donors over a one year period. 
Each patient sample was subdivided into eight aliquots (``tubes") and analyzed with different biomarker combinations, five markers per tube (most markers are proteins). 
In addition to the markers, the forward scatter (FS) and side scatter (SS) of each sample was also measured in each tube. 
Hence, we have $ 359 \times8  = 2,872$ samples and each sample is seven-dimensional
(five markers and the two scatters).
Tube 1 is an isotype control used to detect non-specific antibody binding and Tube 8 is an unstained control for identifying background or autofluorescence of the system. 
Since the data has been compensated for autofluorescence and spectral overlap by experts, 
we omit these tubes from our analysis.
The disease status (AML/healthy) of $23$ AML patients and $156$ healthy donors are provided as training set,  and 
the challenge is to determine the disease status of the rest of the samples,  
$20$ AML and  $157$ healthy, based only on the information in the training set. 
The complete dataset is  available at {\tt http://flowrepository.org/}.

The side scatter (SS) and all of the  fluorescence channels are transformed logarithmically,  
but the forward scatter (FS) is 
linearly transformed to the interval [0,1] so that all channels have values 
in the  same range.
This removes any bias towards FS channel in the multi-dimensional clustering phase.
After preprocessing, an FC  sample is stored as  an $n\times p$ matrix $A$, where 
the element  $A(i,j)$ quantifies the $j^{th}$ feature in the $i^{th}$ cell, and 
$p$ is the number of features measured in each of $n$ cells. 
In this dataset,  $p=7$ for each tube and $n$ varies among the samples.

\subsection{Identifying cell populations in each sample} 
\label{sec:clustering}
We employ a two-stage clustering approach for identifying phenotypically similar cell populations (homogeneous clusters of cells) in each sample.
At first, we apply the $k$-means clustering algorithm for a wide range of values for $k$, and select the optimum number of clusters $k^*$ by simultaneously optimizing the Calinski-Harabasz and S\_Dbw cluster validation methods~\citep{halkidi2001clustering}.
Next, we model the clusters identified by the $k$-means algorithm with a finite mixture model of multivariate normal distributions. 
In the mixture model, the $i^{th}$ cluster is  represented by two distribution parameters $\boldsymbol {\mu_i}$, 
the $p$-dimensional mean vector, and $\Sigma$, the $p\times p$ covariance matrix.
The distribution parameters for each cluster are then estimated using the Expectation-Maximization (EM) algorithm. 
The statistical parameters of a cluster are used to describe the corresponding cell population in the rest of the  analysis. 
 
\subsection{Dissimilarity between samples}
\label{sec:cluster_dist}

We calculate the dissimilarity between a pair of cell populations by the Mahalanobis distance between their distributions. 
Let $c_1(\boldsymbol {\mu_1}, \Sigma_1)$ and $c_2(\boldsymbol {\mu_2}, \Sigma_2)$ be two normally distributed clusters and $\Sigma_p$ be the pooled covariance of $\Sigma_1$ and $\Sigma_2$.
The Mahalanobis distance $d(c_1, c_2)$ between the clusters is computed as follows:
\begin{equation} 
\begin{array}{lcl}
    d(c_1, c_2)  =   \frac 1 2 (\boldsymbol {\mu_1} - \boldsymbol {\mu_2})^\top \Sigma_p^{-1} (\boldsymbol {\mu_1} - \boldsymbol {\mu_2}), \ \ \text{where}\\ \\
    \Sigma_p = {((n_1-1) * \Sigma_1 + (n_2-1) * \Sigma_2))}/  {(n_1+n_2-2)}.  
\end{array}
\label{eq:Mahalanobis}
\end{equation}

We compute the dissimilarity between a pair of samples by optimally matching  (in a graph-theoretic model) similar cell clusters and summing up the dissimilarities of the matched clusters.
In earlier work, we have developed a robust variant of a graph matching algorithm called the Mixed Edge Cover (MEC) algorithm that allows a cluster in one sample to be matched with zero, one, 
 or more clusters in the second sample~\citep{azad2010identifying}. 
The cell population in the first sample could be either  absent, or present, or split into two or more cell populations  
in the  second sample.  
These can happen due to changes in biological conditions or due to artifactual errors in  clustering. 

Consider two FC samples $A$ and $B$ consisting of $k_a$ and $k_b$ cell populations such that $A=\{a_1,a_2, ..., a_{k_a}\}$, and $B=\{b_1,b_2, ..., b_{k_b}\}$ where  $a_i$ is the $i^{th}$ cluster from sample $A$ and $b_j$ is the $j^{th}$ cluster from $B$. 
The mixed edge cover computes a mapping {\tt mec}, of clusters across  $A$ and $B$ such that ${\tt mec}(a_i) \in \mathcal P(B)$ and ${\tt mec}(b_j) \in \mathcal P(A)$, where $ \mathcal P(A)$ ($ \mathcal P(B)$) is the power set of  $A$ ($B$).
When a cluster $a_i$ (or $b_j$) remains unmatched under {\tt mec} , i.e., ${\tt mec}(a_i)=\emptyset$, we set $d(a_i,-)=\lambda$,  where the fixed cost $\lambda$ is a penalty for leaving a vertex unmatched.
We set $\lambda$ to $\sqrt{p}$ so that a pair of clusters get matched only if the average squared deviation across all dimensions is less than one.
The cost of a mixed edge cover {\tt mec} is the sum of the dissimilarities of all pairs of matched clusters and the penalties due to the unmatched clusters. 
A minimum cost  mixed edge cover is a mixed edge cover with the minimum cost.
We use this minimum cost  as the dissimilarity $D(A,B)$ between a pair of samples $A$ and $B$: 
\begin{equation} \label{eq:mec}
\min_{ \substack{ \text{mixed}\ \text{edge}\\ \text{covers}, \ {\tt mec} } } \ 
( \sum_{\substack{1\leq i \leq k_a \\ b_j\in{\tt mec}(a_i) }} d(a_i, b_j) \ +   
\sum_{\substack{1\leq i \leq k_b \\ a_j\in{\tt mec}(b_i) }} d(b_i, a_j) ),
\end{equation}
where $d(a_i, b_j)$ is computed from Equation~(\ref{eq:Mahalanobis}). 
A minimum cost mixed edge cover can be computed by a modified minimum weight perfect matching algorithm
in $O(k^3\log k)$ time where $k$ is the maximum number of clusters in a 
sample \citep{azad2010identifying}. 
The number of cell clusters $k$ is typically small (fewer than
fifty for the AML data), and  the dissimilarity between a pair of samples can be computed in less than a second on a desktop computer. 


\subsection {Creating templates from a collection of samples}
\label{sec:templates}
\begin{figure}[!tpb]
\centerline{\includegraphics[scale=.7]{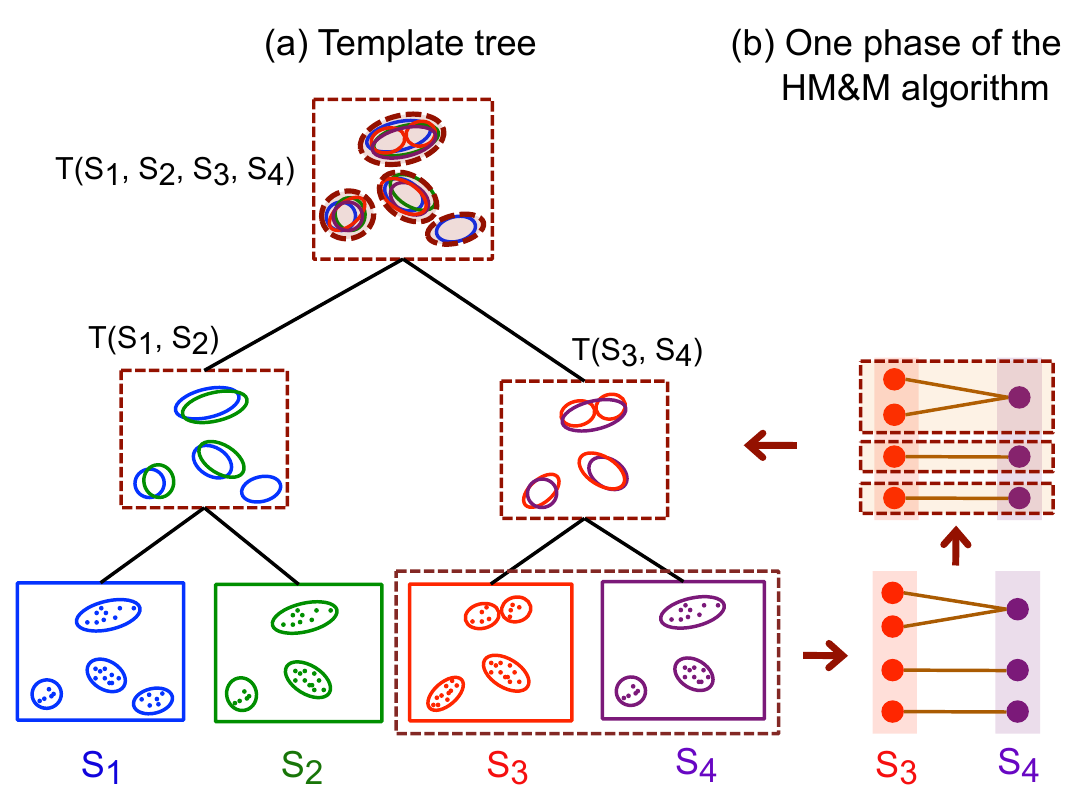}}
\caption{(a) A hierarchical template tree created by the HM\&M algorithm from four hypothetical samples $S_1, S_2, S_3$ and $S_4$. Cells are denoted with dots, and clusters with solid ellipses in the samples are at  the leaves of the tree.
An internal node represents a template created from its children, and the root represents the template of these four samples.
A meta-cluster is a homogeneous collection of clusters and is denoted by a dashed ellipse inside the template. (b) One phase of the HM\&M algorithm creating a sub-template $T(S_3,S_4)$ from samples $S_3$ and $S_4$. At first, corresponding clusters across $S_3$ and $S_4$ are matched by the MEC algorithm, and then the matched clusters are merged to construct new meta-clusters.}
\label{fig:template_example}
\end{figure}

We have designed a hierarchical matching-and-merging (HM\&M) algorithm that arranges a set of similar samples into a binary \emph{template tree} data structure \citep{azad2012matching}.
A node in the tree represents either a sample (leaf node) or a template (internal node).
In both cases, a node is characterized by a finite mixture of multivariate  normal distributions each component of which is a cluster or meta-cluster.
Fig.~\ref{fig:template_example} shows an example of a template-tree created from four hypothetical samples, $S_1, S_2, S_3$, and $S_4$.

Let a node $v_i$ (representing either a sample or a template) in the template tree consist of $k_i$ clusters or meta-clusters $c^i_1$, $c^i_2$, $\ldots$, $c^i_{k_i}$.
A node $v_i$ is called an ``orphan" if it does not have a parent in the template-tree.
Consider $N$ flow cytometry samples $S_1, S_2, \ldots, S_N$ belonging to a  class.
Then the HM\&M algorithm for creating a template tree from these samples can be described 
by  the following three steps. 

1. \emph{Initialization}: Create a node $v_i$ for each of the $N$ samples  $S_i$. 
Initialize all these nodes to the set of orphan nodes. 
 Repeat the matching and merging steps until a single orphan node remains. 

2. \emph{Matching}: Compute the dissimilarity $D(v_i,v_j)$ between every pair of nodes  $v_i$ and $v_j$ in the current ${\tt Orphan}$ set with the mixed edge cover 
algorithm.  (using Equation~(\ref{eq:mec}))

3. \emph{Merging:} Find a pair of orphan nodes $(v_i,v_j)$ with minimum dissimilarity $D(v_i, v_j)$ and merge them to create a new node $v_l$. 
Let ${\tt mec}$ be a function denoting the mapping of clusters from $v_i$ to $v_j$.
That is, if $c_x^i\in v_i$ is matched to $c_y^j\in v_j$, then $c_y^j \in  {\tt mec}(c_x^i)$, where  $1\leq x \leq k_i$ and $1\leq y \leq k_j$.
Create a new meta-cluster $c^l_z$ from each set of matched clusters, $c^l_z = \{c_x^i\cup  {\tt mec}(c^i_x)\}$.
Let $k_l$ be the number of the new meta-clusters created above.
Then the new node $v_l$ is created as a collection of these newly created meta-clusters, i.e., $v_l=\{c^l_1,c^l_2, ..., c^l_{k_l}\}$.
The distribution parameters, $(\mu^l_z, \Sigma^l_z)$, of each of the newly formed meta-clusters $c^l_z$ are estimated by the EM algorithm. 
The height of  $v_l$ is set to $D(v_i, v_j)$.
The node $v_l$ becomes the parent of $v_i$ and $v_j$, 
and the set of orphan nodes is updated by including $v_l$ and   deleting $v_i$ and $v_j$ from it. 
If there are orphan nodes remaining, we return to the matching step, and otherwise, we terminate. 

When the class labels of samples are not known {\em a priori\/}, the roots of well-separated branches of tree give different class templates.
However, if samples belong to the same class -- as is the case for the AML dataset studied in this paper, the root of the template-tree gives the class-template. 
The HM\&M algorithm requires $O(N^2)$ dissimilarity computations and $O(N)$ merge operations for creating a template from a collection of $N$ samples.
Let $k$ be the maximum number of clusters or meta-clusters in any of the nodes of the template-tree.
Then a dissimilarity computation takes $O(k^3\log k)$ time whereas the merge operation takes $O(k)$ time when distribution parameters of the meta-clusters are computed by maximum likelihood estimation. 
Hence, the time complexity of the algorithm is $O(N^2k^3 \log k)$, which is $O(N^2)$ for bounded $k$.
The complexity of the algorithm can be reduced to $O(N \log N)$ by avoiding the computation of 
all pairwise dissimilarities between the samples, for larger numbers of samples $N$,
but we did not need to do this here. 

\subsection{Classification score of a sample in AML dataset}
Consider a sample $X$ consisting of $k$ cell populations $S=\{c_1, c_2, ..., c_k\}$,
with the $i^{th}$ cluster $c_i$ containing $|c_i|$ cells.
Let $T^-$ and $T^+$ be the templates created from AML-negative (healthy) and AML-positive training samples,  respectively.
We now describe how to compute a score $f(X)$ in order to classify the  sample $X$ to 
either the healthy class or the AML class. 

\begin{figure*}[!tbp]
\centerline{\includegraphics[scale=.46]{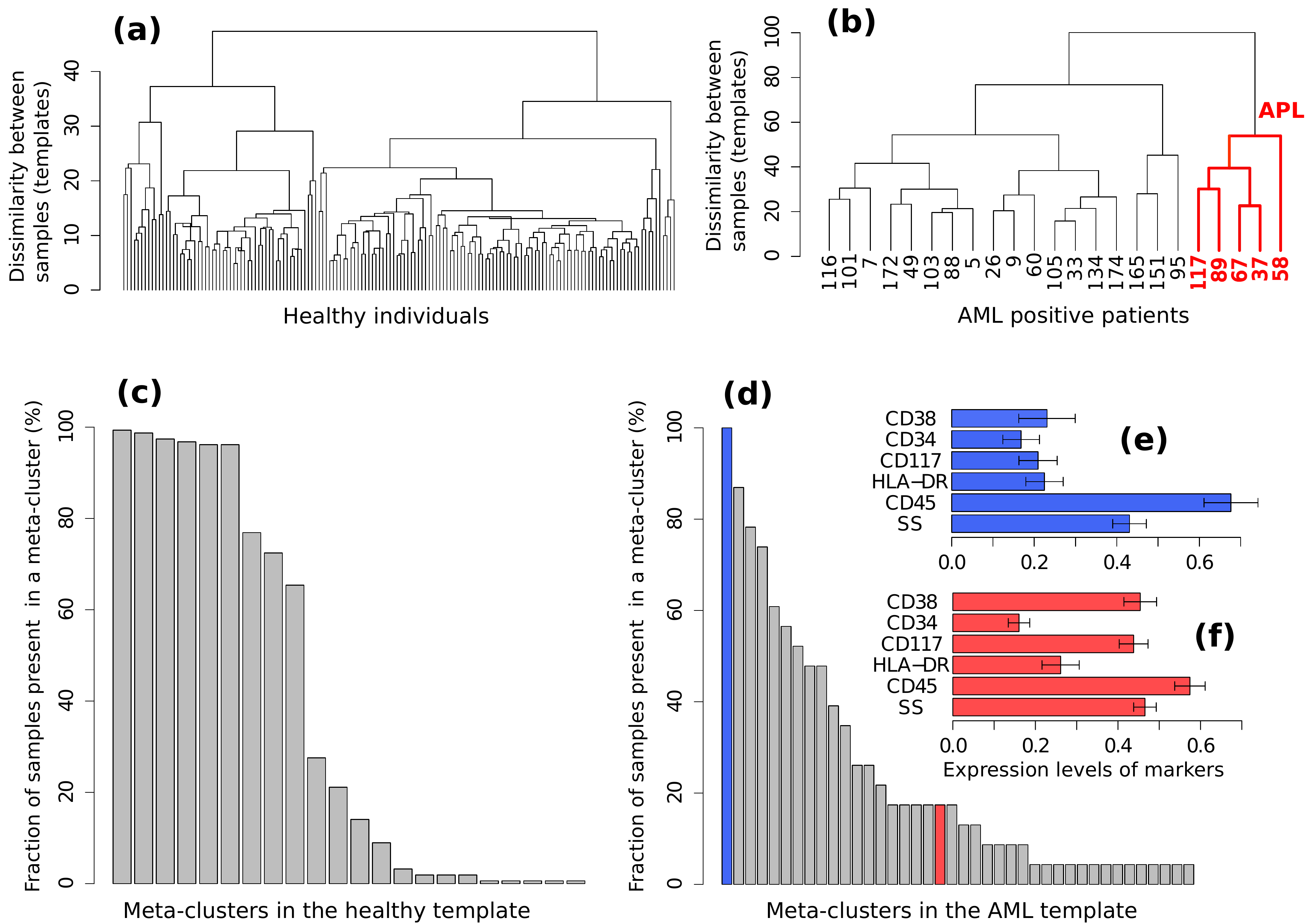}}
\caption{The healthy and AML templates created from Tube 6. 
{\bf (a)} The template-tree created from  156 healthy samples in the training set. 
{\bf (b)} The template-tree created from  23 AML samples in the training set. 
Samples in the red subtree  exhibit  the characteristics of Acute Promyelocytic Leukemia (APL) as shown in Subfigure (f).
{\bf (c)} Fraction of 156 healthy samples present in each of the 22 meta-clusters in the healthy template. 
Nine meta-clusters, each of them shared by at least 60\% of the healthy samples, form the core of the healthy template.
{\bf (d)} Fraction of 23 AML samples present in each of the 40 meta-clusters in the AML template.
The AML samples, unlike the healthy ones, are heterogeneously distributed over the meta-clusters.
{\bf (e)} The expression levels of markers in the meta-cluster shown with blue bar in 
Subfigure (d).
(Each horizontal bar in Subfigures (e) and (f) represents the average expression of a marker and the error bar shows its  standard deviation.) 
This meta-cluster represents lymphocytes denoted by medium SS and high CD45 expression 
and therefore does not express the AML-related markers measured in Tube 6.
{\bf (f)} Expression of markers in a meta-cluster shown with red bar in Subfigure (d).
This meta-cluster denotes myeloblast cells as defined by the SS and CD45 levels.  
This meta-cluster expresses HLA-DR$^-$CD117$^{+}$CD34$^-$CD38$^+$, a   characteristic immunophenotype of APL. 
Five AML samples sharing this meta-cluster are similar to each other as shown in the red 
subtree in Subfigure (b). 
}
\label{fig:templates_tube6}
\end{figure*}

The intuition behind the score is as follows. 
An AML sample contains two kinds of cell populations: 
(1) AML-specific myeloblasts and myeloid cells, and 
(2) AML-unrelated cell populations, such as 
lymphocytes. The former cell populations correspond to the immunophenotypes 
of  AML-specific metaclusters in the AML template,  and hence 
when we compute  a mixed edge cover between the AML template and an AML  sample,
these  clusters get matched to each other. 
(Such clusters in the sample do not match to any metacluster in the healthy template.) 
Hence we assign a positive score to a cluster in sample when it satisfies this condition,
signifying that it is indicative of AML. 
AML-unrelated cell populations in a sample could match to meta-clusters in the healthy template,
and also to AML-unrelated meta-clusters in the AML template. 
When either of these conditions is satisfied, a cluster gets a negative score, 
signifying that it is not indicative of AML. 
Since AML affects only the myeloid cell line and its progenitors, it affects only a
small number of AML-specific cell populations in an AML sample. 
Furthermore, different subtypes of AML affect different cell types in the myeloid cell line. 
Hence  there are many more clusters common to healthy samples than there are 
AML-specific clusters common to AML samples. 
(This is illustrated later in Fig.~\ref{fig:templates_tube6} (c) and (d).)
Thus we make the range of positive scores  relatively higher  than 
the range of negative scores. 
This scoring system is designed to reduce the possibility of a false negative 
(an undetected AML-positive patient), since this is more serious in the diagnosis of 
AML. Additional data such as chromosomal translocations and images of bone marrow from 
microscopy could confirm an initial  diagnosis of AML from flow cytometry. 

In the light of the discussion above, we need to  identify AML-specific metaclusters initially.
Given the templates $T^+$ and $T^-$, we create a complete bipartite graph 
with the meta-clusters in each template as vertices, and with each  edge  weighted by 
the Mahalanobis distance between its endpoints. 
When we compute a minimum cost mixed edge cover in this graph, we will match meta-clusters 
common to both templates, and such  meta-clusters represent non-myeloid cell populations
that are not AML-specific. On the other hand, meta-clusters in the AML template $T^+$ 
that are not matched to a meta-cluster in the healthy template $T^-$
correspond to AML-specific metaclusters. 
We denote such meta-clusters in the AML template $T^+$ by the set $M^+$. 

Now we can proceed to compare a sample against the template for healthy samples and 
the template for AML. 
We compute a minimum cost mixed edge cover between  a  sample $X$ and the healthy template 
$T^-$, and let ${\tt mec^-}(c_i)$ denote the set of meta-clusters in $T^-$ mapped to 
a cluster $c_i$ in the sample $X$. 
Similarly, compute a minimum cost mixed edge cover between $X$ and 
the AML template $T^+$, and let ${\tt mec^+}(c_i)$ denote the set of meta-clusters in $T^+$ 
mapped to a  cluster $c_i$. These sets could be empty if $c_i$ is unmatched 
in the mixed edge cover. 
We compute the average Mahalanobis distance between $c_i$ and the meta-clusters matched to 
it in the template $T^-$, and define this as the dissimilarity 
$d(c_i, {\tt mec^-}(c_i))$. 
From the formulation of the mixed edge cover in~\citep{azad2010identifying},  
we have $d(c_i, {\tt mec^-}(c_i)) \leq 2\lambda$. 
Hence we define the {\em similarity\/} between 
$c_i$ and ${\tt mec^-}(c_i)$ as  $s(c_i, {\tt mec^-}(c_i)) = 2\lambda-d(c_i, {\tt mec^-}(c_i))$.
By analogous reasoning, the similarity between $c_i$ and ${\tt mec^+}(c_i)$ is defined as 
$s(c_i, {\tt mec^+}(c_i)) = 2\lambda-d(c_i, {\tt mec^+}(c_i))$.

The score of a sample is the sum of the scores of its clusters. 
We define the score of a cluster $c_i$,  $f(c_i)$,  
as the sum of two functions $f^+(c_i)$ and $f^-(c_i)$ multiplied with suitable weights. 
A positive score indicates that the sample 
belongs to AML, and a negative score indicates that it is healthy. 

The function  $f^+(c_i)$ contributes  a positive score to the sum if $c_i$ is 
matched to an AML-specific meta-cluster 
in the mixed edge cover between the sample $X$ and the AML template $T^+$,
and a non-positive score otherwise. For the latter case,  there  are two subcases: 
If $c_i$ is unmatched in the mixed edge cover, it corresponds to none of the meta-clusters 
in the template $T^+$, and we assign it a zero score. 
If $c_i$ is matched only to non-AML specific meta-clusters in the AML template $T^+$, 
then we assign it a small negative score to 
indicate that it likely belongs to the healthy class 
(recall that $k$ is the number of clusters in sample $X$). 
Hence 
\begin{eqnarray*}
f^+(c_i) &=&   \begin{cases}
                   s\left(c_i, {\tt mec^+}(c_i)\right), &\text{if\ \ } {\tt mec^+}(c_i) \cap M^+ \neq  \emptyset,\\
     -\frac{1}{k} \left[s(c_i, {\tt mec^+}(c_i))\right], &\text{if\ \ } {\tt mec^+}(c_i) \cap M^+ =  \emptyset,\\
                   & {\text{and\  \ } \tt mec^+}(c_i) \neq  \emptyset,\\
                   0,                              &\text{if\ \ } {\tt mec^+}(c_i) = \emptyset.\\
\end{cases}
\end{eqnarray*}

The function $f^-(c_i)$ contributes a negative score to a cluster $c_i$ in the sample $X$ 
if it is matched with some meta-cluster in the healthy template $T^-$,
indicating that it likely belongs to the healthy class. 
If it is not matched to any  meta-cluster in $T^-$, then we assign it a positive score 
$\lambda$. This latter subcase accounts for AML-specific clusters in the sample,
or a cluster that is in neither template. In this last case, we 
acknowledge the diversity of cell populations  in AML samples. 
Hence we have 
\begin{eqnarray*} 
f^-(c_i) &=& \begin{cases}
             -\frac{1}{k} \left[s(c_i, {\tt mec^-}(c_i))\right], &\text{if\ \ } {\tt mec^-}(c_i) \neq \emptyset,\\
\lambda, &\text{if\ \ } {\tt mec^-}(c_i) = \emptyset.\\
\end{cases}   \\ 
\end{eqnarray*}

Finally, we define 
\begin{equation} 
f(X) = \sum_{c_i\in X} \frac{|c_i|}{|X|} \frac{1}{2} (f^+(c_i) + f^-(c_i)).
\label{eq:classif_score}
\end{equation}
Here $|X|$ is the number of cells in the sample $X$. The score of a cluster $c_i$ is weighted by
the fractional abundance of cells in it. 

\vspace{-0.2in}

\begin{figure*}[!tpb]
\centerline{\includegraphics[scale=.4]{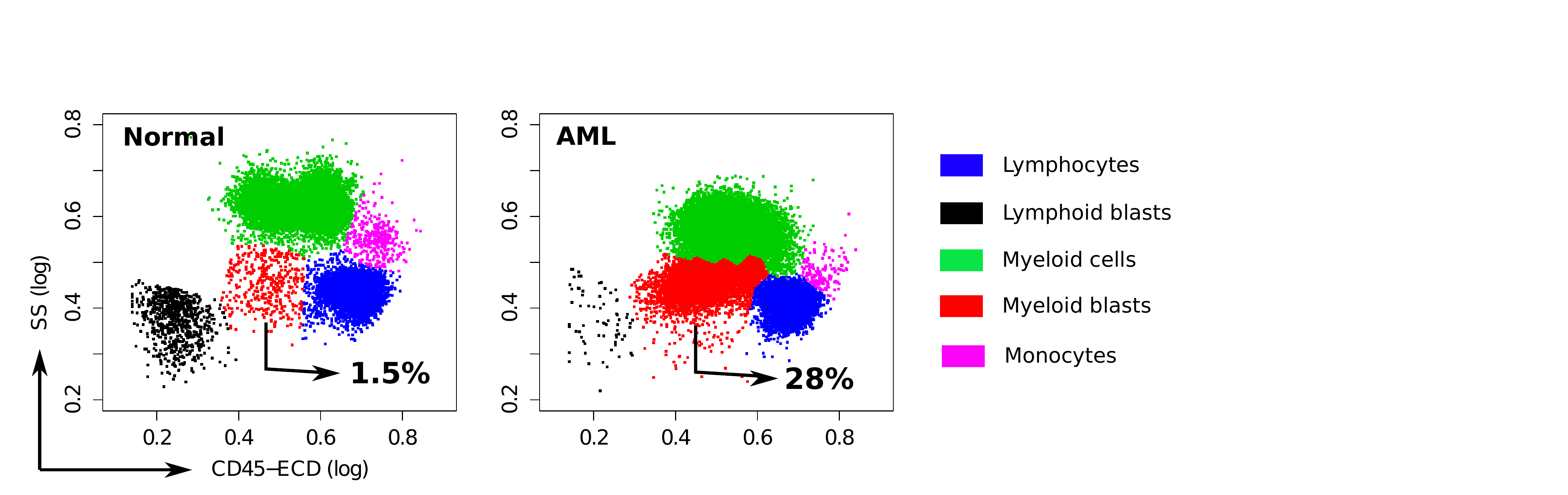}}
\caption{Cell types identified on the side scatter (SS) and CD45 channels for a healthy and an AML positive sample.
Cell populations are discovered in the seven-dimensional samples with the clustering algorithm and then projected on these channels for visualization.
A pair of clusters denoting the same cell type is marked with the same color.
The proportion of myeloid blast cells (shown in red) increases significantly in the AML sample.
 }
   \label{fig:AML_blasts}
\end{figure*}

\section{Results}
\subsection{Cell populations in healthy and AML samples}
In each tube, we identify cell populations in the samples using  the clustering algorithm described in Section~\ref{sec:clustering}.
Each sample contains five major cell types that can be seen when cell clusters are projected on the side scatter (SS) and CD45 channels, as depicted in  Fig.~\ref{fig:AML_blasts}.
(Blast cells are immature progenitors of myeloid cells or lymphocytes.) 
The side scatter measures the granularity of cells, whereas CD45 is variably expressed by different white blood cells (leukocytes).
AML is initially diagnosed by  rapid growth of immature myeloid blast cells with medium SS and CD45 expressions~\citep{lacombe1997flow} marked in red in Fig.~\ref{fig:AML_blasts}.
According to the WHO guidelines, AML is initially confirmed when the sample contains more than 20\% blasts.
This is the  case for all, except one of the  AML samples in the DREAM6/FlowCAP2  training set, 
and the latter  will be discussed later. 


\subsection{Healthy and AML templates} 
From  each tube of the AML dataset, using the training samples,
we build two templates: one  for healthy samples, and one for AML. 
As described in Section~\ref{sec:templates},  the HM\&M algorithm organizes samples of the same class into a binary template tree whose root represents the class template.
The template trees created from the healthy and AML training samples in Tube 6 
are shown in Subfigures \ref{fig:templates_tube6}(a) and \ref{fig:templates_tube6}(b) respectively.
The height of an internal node in the template tree measures the dissimilarity between its left and right children, whereas the horizontal placement of a sample is arbitrary.
In these trees, we observe twice as much heterogeneity in the AML samples than  among the healthy samples (in the dissimilarity measure), 
despite the number of healthy samples being five times as numerous as the  AML samples.
The larger heterogeneity among AML samples is observed in other tubes as well.
The  template-tree for AML partitions these samples into different subtrees that possibly  denote different subtypes of AML.
For example, the subtree in Fig.~\ref{fig:templates_tube6}(b) that is colored red 
includes samples (with subject ids 37, 58, 67, 89, and 117) with immunophenotypes of 
Acute Promyelocytic Leukemia (APL) (discussed later in this section).

Together, the meta-clusters in a healthy template  represent a healthy immune profile in the feature space of a tube from which the template is created. 
We obtained $22$ meta-clusters in the healthy template created from Tube 6.
The percentage of  samples from the training set participating in each of these meta-clusters is shown in Fig.~\ref{fig:templates_tube6}(c).
Observe that $60\%$ or more of the healthy samples participate in the nine most 
common meta-clusters (these constitute  the core of the healthy template).
The remaining thirteen meta-clusters include populations from a small fraction of samples. 
These populations could correspond to biological variability  in the healthy samples,
variations in the FC experimental protocols,  
and possibly also from the splitting of populations 
that could be an artifact of the clustering algorithm. 

The AML template created from Tube 6 consists of forty  meta-clusters
(almost twice the number in the more numerous healthy samples).
Fig.~\ref{fig:templates_tube6}(d) shows that, unlike the healthy samples, the AML samples are 
heterogeneous with respect to  the meta-clusters they participate in:  
There are  $21$ meta-clusters that include cell populations from at least $20\%$ of the 
AML samples.
Some of the  meta-clusters common to a large number of AML samples  represent 
non-AML specific cell populations.
For example, Fig.~\ref{fig:templates_tube6}(e) shows the average marker expressions of the meta-cluster shown in the blue bar in Fig.~\ref{fig:templates_tube6}(d).
This meta-cluster has low to medium side scatter and high CD45 expression, and therefore represents lymphocytes  (Fig.~\ref{fig:AML_blasts}).
Since lymphocytes are not affected by AML, this meta-cluster does not express any AML-related markers, and hence can be described as HLA-DR$^-$CD117$^-$CD34$^-$CD38$^-$,  as expected.
Fig.~\ref{fig:templates_tube6}(f) shows the expression profile of another meta-cluster 
shown in the red bar in Fig.~\ref{fig:templates_tube6}(d).
This meta-cluster consists of five cell populations from five AML samples (with subject ids  37, 58, 67, 89, and 117) and exhibits medium side scatter and CD45 expression and therefore, 
represents myeloid blast cells.
Furthermore, this meta-cluster is HLA-DR$^-$CD117$^+$CD34$^-$CD38$^+$, 
and represents a profile known to be that of Acute Promyelocytic Leukemia 
(APL)~\citep{paietta2003expression}.
APL is subtype M3 in the FAB classification of AML~\citep{bennett1985proposed}) and is characterized by chromosomal translocation of retinoic acid receptor-alpha (RAR$\alpha$) gene on chromosome 17 with the promyelocytic leukemia gene (PML) on chromosome 15, a translocation denoted as t(15;17). 
In the feature space of Tube 6, these APL samples are similar to each other while significantly different from the other  AML samples.
Our template-based classification algorithm groups these samples together in the subtree colored red in the AML template tree shown in Fig.~\ref{fig:templates_tube6}(b). 

\vspace{-0.2in}

\subsection{Identifying meta-clusters symptomatic of AML}  
\label{sec:metaclusters}
In each tube, we register meta-clusters across the AML and healthy templates using the mixed edge cover (MEC) algorithm.
Meta-clusters in the AML template that are not matched to any meta-clusters in the 
healthy template represent  the abnormal, AML-specific immunophenotypes while the matched meta-clusters represent healthy  or non-AML-relevant cell populations.         
Table~\ref{tab:unmatched_mc} lists several unmatched meta-clusters indicative of AML from different tubes. 
As expected, every unmatched meta-cluster displays medium side scatter and CD45 expression 
characteristic of myeloid blast cells, and 
therefore we omit FS, SS, and CD45 values in Table~\ref{tab:unmatched_mc}.
We briefly discuss the immunophenotypes represented by each AML-specific meta-cluster in each tube,
omitting the isotype control Tube 1 and unstained Tube 8. 
 
\begin{table}[!tbp]
\processtable{Some of the meta-clusters characteristic of AML 
for the 23 AML samples in the training set.
In the second column, `$-$',  `low', and  `$+$' denote very low, low and high,
abundance of a marker, respectively, and $\pm$ denotes a marker that is positively expressed by some samples 
and  negatively expressed by others. 
The number of samples participating in a meta-cluster is shown in the third column.
The average fraction of cells 
in a sample participating in a  meta-cluster, and the standard deviation,  
are shown in the fourth column.
\label{tab:unmatched_mc}}
{\begin{tabular}{cccc}\toprule
Tube   & Marker expression & \#Samples & Fraction of cells \\ \toprule
	2  \ \  & Kappa$^\text{low}$Lambda$^\text{low}$CD19$^+$CD20$^-$ & 5 &  $6.3\%(\pm 6.8)$ \\ 
	3  \ \  & CD7$^+$CD4$^-$CD8$^-$CD2$^-$ & 4 &  $18.0\%(\pm 4.8)$ \\ 
	4  \ \  & CD15$^-$CD13$^+$CD16$^-$CD56$^-$ & 17 &  $16.6\%(\pm 6.9)$ \\ 
	4  \ \  & CD15$^-$CD13$^+$CD16$^-$CD56$^+$ & 8 &  $11.1\%(\pm 5.7)$ \\ 
	5  \ \  & CD14$^-$CD11c$^-$CD64$^-$CD33$^+$ & 10 &  $13.5\%(\pm 5.2)$ \\ 
	5 \ \  & CD14$^-$CD11c$^+$CD64$^-$CD33$^+$ & 18 &  $10.8\%(\pm 3.8)$ \\ 
	5  \ \  & CD14$^\text{low}$CD11c$^+$CD64$^\text{low}$CD33$^+$ & 6 &  $13.8\%(\pm 4.3)$ \\ 
	6  \ \  & HLA-DR$^+$CD117$^+$CD34$^+$CD38$^+$ & 11 &  $13.3\%(\pm 2.6)$ \\ 
	6  \ \  & HLA-DR$^+$CD117$^{\pm}$CD34$^+$CD38$^+$ & 13 &  $17.3\%(\pm 6.6)$ \\ 
	6  \ \  & HLA-DR$^-$CD117$^{\pm}$CD34$^-$CD38$^+$ & 5 &  $12.9\%(\pm 4.7)$ \\ 
	7  \ \  & CD5$^-$CD19$^+$CD3$^-$CD10$^-$ & 3 &  $12.3\%(\pm 2.4)$ \\ 
	7  \ \  & CD5$^+$CD19$^-$CD3$^-$CD10$^-$ & 3 &  $10.0\%(\pm 8.5)$ \\ 
	7  \ \  & CD5$^-$CD19$^-$CD3$^-$CD10$^+$ & 1 &  $9.9\%$ \\ 
     	 \botrule
\end{tabular}}{}
\vspace{-.3in}
\end{table}

\begin{figure*}[!tpb]
\centerline{\includegraphics[scale=.35]{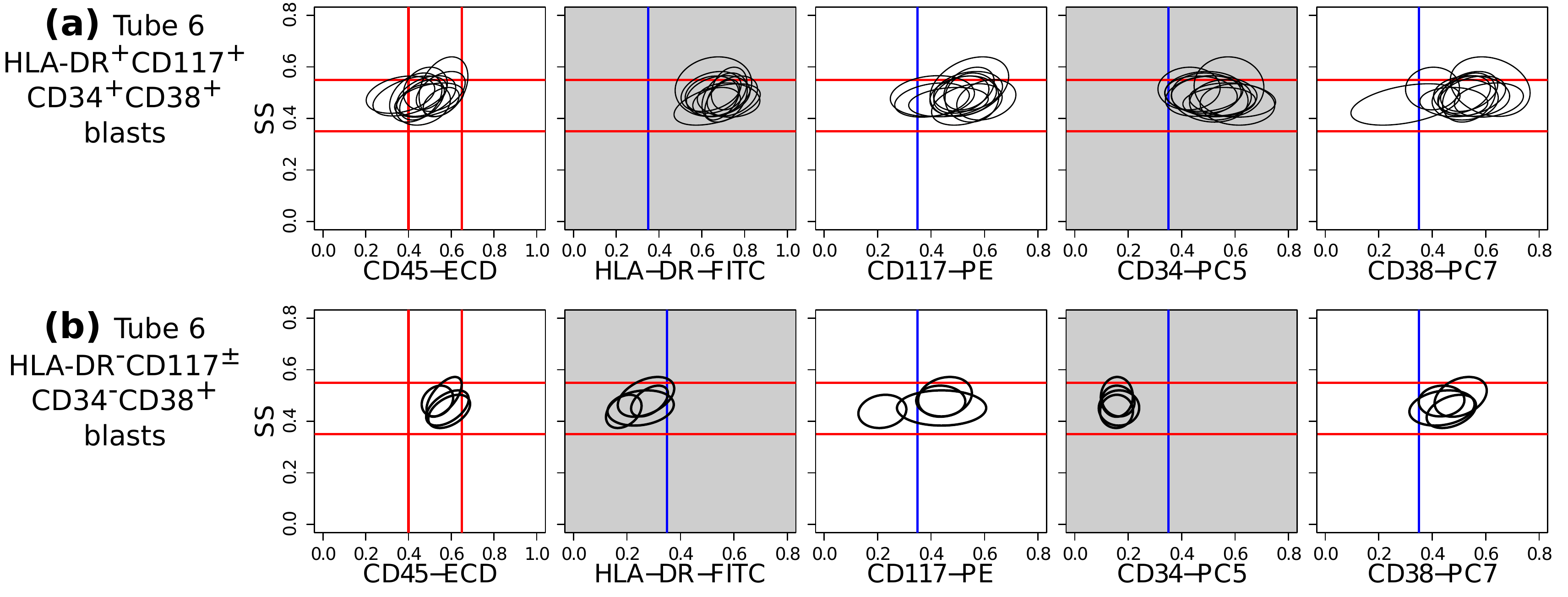}}
\caption{Bivariate contour plots (side scatter vs. individual marker) for two meta-clusters (one in each row) indicative of AML.
The ellipses in a subplot denote the 95th quantile contour lines of cell populations included in the corresponding meta-cluster.
Myeloblast cells have medium side scatter (SS) and CD45 expressions.
The red lines indicate approximate myeloblast boundaries (located on the left-most subfigures in each row and extended horizontally to the  subfigures on the right) and confirm that these meta-clusters represent immunophenotypes of myeloblast cells. 
Blue vertical lines denote the +/- boundaries of a marker. 
Gray subplots show contour plots of dominant markers defining the meta-cluster in the same row. 
{\bf (a)} HLA-DR$^+$CD117$^+$CD34$^+$CD38$^+$ meta-cluster shared by 11 AML samples in Tube 6.
{\bf (b)}  HLA-DR$^-$CD117$^{\pm}$CD34$^-$CD38$^+$ meta-cluster shared by 5 AML samples in Tube 6. This meta-cluster is indicative of acute promyelocytic leukemia (APL).
These bivariate plots are shown for illustration only, 
since the populations of specific cell types are identified from seven-dimensional data. 
}\label{fig:mc}
\end{figure*}

{\bf Tube 6} is the most important panel for diagnosing AML since it includes several markers expressed by AML blasts.
HLA-DR is an MHC class II cell surface receptor complex that is expressed on 
antigen-presenting cells, e.g., B cells, dendritic cells, macrophages, and activated T cells.
It is expressed by myeloblasts in most subtypes of AML except M3 and M7~\citep{campana2000immunophenotyping}. 
CD117 is a tyrosine kinase receptor (c-KIT) expressed in blasts of some cases 
($30-100\%$) of AML~\citep{campana2000immunophenotyping}.
CD34 is a cell adhesion molecule expressed on different stem cells and on the blast cells of many cases of AML ($40\% $) \citep{mason2006immunophenotype}.
CD38 is a glycoprotein found on the surface of blasts of several subtypes of AML but  usually not expressed in the 
M3 subtypes of AML~\citep{keyhani2000increased}.
In Tube 6, we have identified two meta-clusters with high expressions of HLA-DR and CD34.
One of them also expresses CD117 and CD34, and Fig.~\ref{fig:mc}(c) shows the bivariate contour plots of the cell populations contained in this meta-cluster.
The second meta-cluster expresses positive but low levels of CD117 and CD34.
These two HLA-DR$^+$CD34$^+$ meta-clusters together 
are present in  $18$ out of the $23$  training AML samples.
The remaining five samples (subject id: 5, 7, 103, 165, 174) express HLA-DR$^-$CD117$^{\pm}$CD34$^-$CD38$^+$ myeloblasts, which is an immunophenotype of APL~\citep{paietta2003expression} as was discussed earlier.
Fig.~\ref{fig:mc}(d) shows the bivariate contour plots of this APL-specific meta-cluster.

{\bf Tube 5} contains several antigens typically expressed by AML blasts, of which CD33 is the most important. 
CD33 is a transmembrane receptor protein usually expressed on immature myeloid cells of the majority of cases of AML ($91\%$  reported in~\citep{legrand2000immunophenotype}).
The AML specific meta-clusters identified from markers in Tube 5 
(see Table~\ref{tab:unmatched_mc}) include CD33$^+$ myeloblasts from 
every sample in the training set.
Several of the CD33$^+$ populations also express CD11c, a type I transmembrane protein found on monocytes, macrophages and neutrophils.
CD11c is usually expressed by blast cells in acute myelomonocytic leukemia (M4 subclass of AML),
and acute monocytic leukemia (M5 subclass of AML)~\citep{campana2000immunophenotyping}.
Therefore CD14$^-$CD11c$^+$CD64$^-$CD33$^+$ meta-cluster could represent patients with M4 and M5 subclasses of AML.
We show the bivariate contour plots of this meta-cluster in Fig.~\ref{fig:mc}(b) .

{\bf Tube 4} includes several markers usually expressed by AML blasts, of which CD13 is the most important. 
CD13 is a zinc-metalloproteinase enzyme that binds to the cell membrane and degrades regulatory peptides~\citep{mason2006immunophenotype}.
CD13 is expressed on the blast cells of the majority of cases of AML (95\% as reported in \citep{legrand2000immunophenotype}).
Table~\ref{tab:unmatched_mc} shows two AML-specific  meta-clusters  detected from the blast cells 
in Tube 4.
In addition to CD13, eight AML samples express CD56 glycoprotein that is naturally expressed on NK cells, a subset of CD4$^+$ T cells and a subset of CD8$^+$ T cells. 
Raspadori et al.~\citep{raspadori2001cd56} reported that CD56 was more often expressed by myeloblasts in FAB subclasses  M2 and M5, which covers about 42\% of AML cases in a study by Legrand et al.~\citep{legrand2000immunophenotype}.
In this dataset, we observe more AML samples expressing CD13$^+$CD56$^-$ blasts than expressing CD13$^+$CD56$^+$ blasts, which conforms to the findings of Raspadori et al.~\citep{raspadori2001cd56}.
Fig.~\ref{fig:mc}(a) shows the bivariate contour plots of the CD13$^+$CD56$^-$ meta-cluster.

{\bf Tube 2} is a B cell panel measuring B cell markers CD19 and CD20,  
and Kappa ($\kappa$) and Lambda ($\lambda$), immunoglobulin light chains present on the surface of antibodies produced by B lymphocytes.
B-cell specific markers are occasionally co-expressed with myeloid antigens especially in FAB M2 subtype of AML (with chromosomal translocation t(8;21)) \citep{campana2000immunophenotyping, walter2010aberrant}.
In Tube 2, we have identified a meta-cluster in the myeloblasts that  expresses high levels of CD19 and low levels of Kappa and Lambda.
The five samples with subject ids  5, 7, 103, 165, and 174 participating in this meta-cluster possibly belong to the FAB-M2 subtype of AML.
{\bf Tube 3} is a T cell panel measuring T cell specific markers CD4, CD8, CD2, and CD7.
{\bf Tube 7} is a lymphocyte panel with several markers expressed on T and B lymphocytes and is less important in detecting AML since they are infrequently expressed by AML blasts.

\begin{table*}[!t]
   \centering
\processtable{Four statistical measures  evaluating the performance of the template-based classification in the training set and test set of the AML data.
The statistical measures are computed for each tube separately and two combinations of tubes.  
\label{tab:classification_stats}}
{\begin{tabular}{c cccc p{.5cm}cccc}\toprule
 \ \ \ \ \ {\bf Tubes}\ \ \ \   & \multicolumn{4}{c}{\bf Training set} & & \multicolumn{4}{c}{\bf Test set} \\ 
\cmidrule(lr){2-5} \cmidrule(lr){7-10}
   		& Precision & Recall & Specificity & F-value && Precision & Recall              & Specificity & F-value\\ \toprule
4 		&  	 0.94  & 	0.74  	& 	0.99  	& 	0.83 	&&	1.00      &     0.75     &  	1.00    &  	0.86\\
5 		&  	 0.75  & 	0.91  	& 	0.96  	& 	0.82 	&&	0.65      &     0.85     &   	0.94    & 	0.74\\
6 		&  	 1.00  & 	0.70  	& 	1.00  	& 	0.82 	&&	1.00      &     0.80     &   	1.00    & 	0.89\\
All (2-7)     &  	 1.00  & 	0.74  	& 	1.00 	& 	0.85	&&	1.00      &     0.85     &   	1.00    & 	0.92 \\
4,5,6 	& 	 1.00  & 	0.96  	& 	1.00  	& 	0.98	&&	1.00      &     1.00     &   	1.00    &	1.00 \\
     	 \botrule
\end{tabular}}{}
\end{table*}

\begin{figure*}[!tpb]
\centerline{\includegraphics[scale=.5]{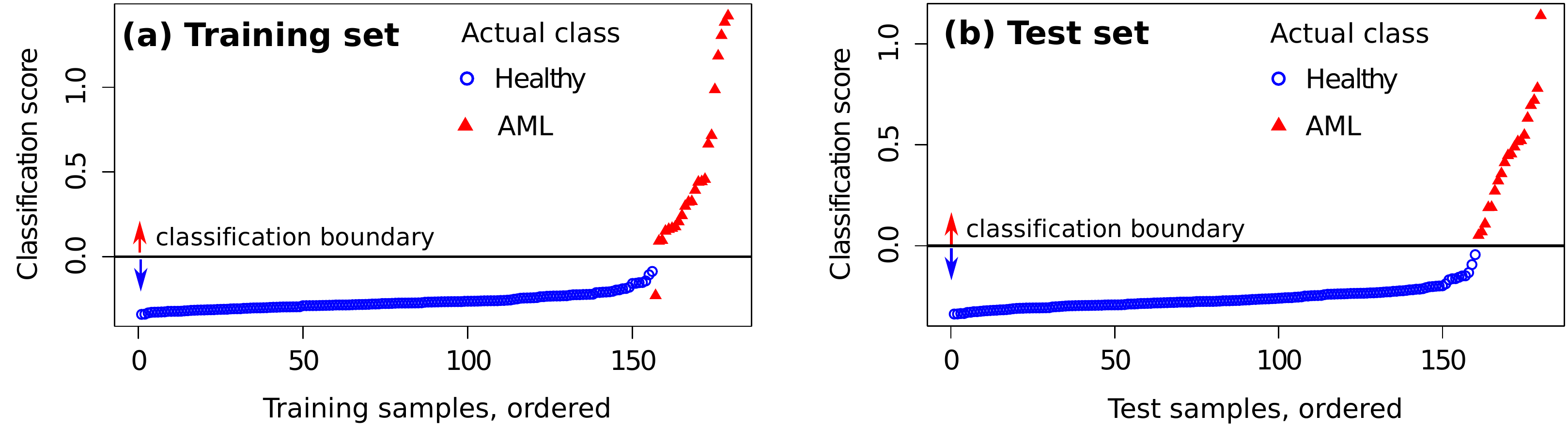}}
\caption{Average classification score from Tubes 4,5,6 for each sample in the (a) training set and (b) test set. 
Samples with scores above the horizontal line are classified as  AML, and as  healthy otherwise.
The actual class of each sample is also shown. 
An AML sample (subject id 116) is always misclassified in the training set, and this is discussed
in the text.
}
\label{fig:classificationTrainTest}
\end{figure*}

\vspace{-0.2in}

\subsection{Impact of each tube in the classification}
As discussed in the methods section, we build six independent classifiers based on the healthy and AML templates created from Tubes 2-7 of the AML dataset. 
A sample is classified as an AML sample if the classification score is positive, and as a healthy sample otherwise. 
Let true positives (TP) be the number of AML samples correctly classified, true negatives (TN) be the number of healthy samples correctly classified, false positives (FP) be the number of healthy samples incorrectly classified as AML, and false negatives (FN) be the number of AML samples incorrectly classified as healthy.
Then, we evaluate the performance of each template-based classifier with 
the well-known four statistical measures: Precision, Recall(Sensitivity), Specificity, and 
F-value, defined as 
$\text{Precision}=\frac{\text{TP}}{\text{TP}+\text{FP}}$,  $\text{Recall(Sensitivity)}=\frac{\text{TP}}{\text{TP}+\text{FN}}$,
$\text{Specificity}=\frac{\text{TN}}{\text{FP}+\text{TN}}$, and  
$\text{F-value}= \frac{2(\text{Precision} \times \text{Recall})}{\text{Precision} + \text{Recall}}$.
These four measures take values in the interval [0,1], and the higher the values the better the classifier.

First, we evaluate the impact of each tube in the classification of the training samples.
For a training sample $X$, the classification score is computed by comparing it 
with the healthy and AML templates created from the training set after removing $X$.
The predicted status of $X$ is then compared against true status to evaluate the classification accuracy.
Table \ref{tab:classification_stats} (left panel) shows various statistical 
measures for  the classifiers defined in Tubes~2-7 of the training set.
The classifiers based on Tubes 4, 5, and 6 have the highest sensitivity because  these tubes include several markers relevant to AML diagnosis~\citep{campana2000immunophenotyping, paietta2003expression}.
The number of true negatives TN is high in every tube since the identification of 
healthy samples does not depend on the detection of AML-specific markers. 
Hence specificity is close to one for all tubes.
Analogously, FP is low for most tubes, and we observe high precision for most tubes. 
The F-value is a harmonic mean of precision and recall,  and denotes the superior 
classification ability of markers in Tubes 4-6.
Averaging scores from all tubes does not improve the sensitivity and F-value dramatically.
However, combining Tubes 4-6 gives almost perfect classification  
with one misclassification for the training set.  
We plot the average classification scores from Tubes 4-6 for the training samples in Fig.~\ref{fig:classificationTrainTest}(a).
The class labels of samples are also shown (blue circles for healthy and red triangles for AML samples).

In Fig.~\ref{fig:classificationTrainTest}(a), we observe an AML sample (subject id 116) with score below the classification boundary.
In this subject, the proportion of myeloid blasts is 4.4\%, which is lower than the minimum 20\% AML blasts necessary to recognize a patient to be AML-positive according to the WHO guidelines~\citep{estey2006acute} (the FAB threshold is even higher,  at $30\%$). 
Hence this is either a rare case of AML,  or  one with minimal residual disease after therapy,
or perhaps it was incorrectly labeled as AML  in the training set. 
Subject 116 was classified with the healthy samples by methods in other published 
work~\citep{biehl2013analysis}.


\subsection{Classifying test samples}
Now we turn to the test samples. 
For each tube, we compute the classification score for each sample in the test set using templates created from the training set and applying Eq.~\ref{eq:classif_score}.
Since the average classification score from Tubes 4-6 performs best for the training set, we use it as a classifier for the test set as well.
Since the status of  test samples was  released after the DREAM6/FlowCAP2 challenge, 
we can determine the classification accuracy of the test samples. 
Fig.~\ref{fig:classificationTrainTest}(b) shows the classification scores of the test samples, where samples are placed in ascending order of classification scores.
In Fig.~\ref{fig:classificationTrainTest}(b), we observe  perfect classification in the test set.
Similar to the training set, we tabulate statistical measures for  the classifiers  in Table~\ref{tab:classification_stats}.


When classifying a sample $X$, we assume the null hypothesis: $X$ is healthy (non-leukemic).
The sample $X$ receives a positive score  if it contains  AML-specific immunophenotypes, 
and  the higher the score,  the stronger the evidence against the null hypothesis.
Since Tube 1 (isotype control) does not include any AML-specific markers, 
it can provide a background distribution for the  classification scores. 
In Tube 1,  174 out of 179 training samples have negative classification scores, but  
five samples have positive scores, with values  less than $0.2$. 
In the best classifier designed from Tubes 4, 5, 6, 
we observe that two AML-positive samples in the training set and three AML-positive samples in the test set have scores between 0 and 0.2. 
The classifier is relatively less confident about these samples;   
nevertheless, the p-values of these five samples (computed from the distribution in Tube 1) 
are still small ($<0.05$),  so that they can be classified as AML-positive. 
The rest of the AML samples in the training and test sets have scores greater than $0.2$ 
and the classifier is quite confident about their status (p-value zero).

Four AML samples in the test set (ids 239, 262, 285, and 326) were subclassified as APL 
by comparing against distinct  template trees for APL and the other AML  samples in the 
training set (cf. Fig.~\ref{fig:templates_tube6} (b)). 

Finally, we state the computational times required on an iMac with four 2.7 GHz cores
and 8 GB memory. Our code is in R. 
Consider a single tube with $359$ samples in it. 
The $k$-means clustering of all samples took one hour, primarily because we need to 
run the algorithm multiple times (about ten on the average) to  find the optimal value of the 
number of clusters. 
Creating the healthy template from $156$ samples in the training set required 
$10$ seconds (s)  on one core, 
and the AML template for $23$ AML samples took $0.5$s on one core. 
Cross validation (leave one out) of the training set took $30$ minutes, 
and computing the classification score for the $180$ test samples took $15$s, 
both  on four cores. We could have reduced  the running time by executing the code in parallel
on more cores. 
We have made the dominant step, the $k$-means clustering of all the samples
with an optimal number of clusters, faster using a GPU, reducing the total 
time to a few minutes. 

\vspace{-0.2in}

\section{Conclusions}

We have demonstrated that an algorithmic pipeline for 
template-based classification can successfully
identify  immunophenotypes of clinical interest in AML. 
These could be used to differentiate the subtypes of AML, which is advantageous 
since prognosis and treatment depends on the subtype. 
The templates enable us to classify AML samples in spite of their  heterogeneity. 
This was accomplished by creating a scoring function that accounts for the 
subtleties in cell populations within AML samples.
We are currently applying  this approach to a larger AML data set,  
and intend to analyze  other heterogeneous data sets.  

\vspace{-0.2in}

\section*{Acknowledgments}
This research was supported by NIH grant IR21EB015707-01, 
NSF grant CCF-1218916, and DOE grant 13SC-003242. 
\vspace{-0.2in}

%
%

\bibliographystyle{natbib}
\bibliography{AML}

%
%
%
%
%
%
%

\end{document}